\documentclass[aps,prl,nofootinbib,twocolumn,preprintnumbers,floatfix,superscriptaddress,longbibliography]{revtex4-1}
\usepackage[colorlinks, linkcolor=red, anchorcolor=blue, citecolor=blue, colorlinks=true, urlcolor=blue]{hyperref}
\usepackage{amsmath}
\usepackage{graphicx}
\usepackage{ulem}
\usepackage{bm}
\usepackage{xspace}
\usepackage{amssymb}
\usepackage{gensymb}
\usepackage{multirow}
\usepackage{threeparttable}
\usepackage{amsthm}
\usepackage{mathrsfs}
\usepackage{indentfirst}
\usepackage{subfigure}
\usepackage[figuresright]{rotating}
\usepackage{epstopdf}
\usepackage{breakurl}
\usepackage{lipsum}
\usepackage{cancel}
\usepackage{slashed}

\begin{document}

\title{Mass Mixing between QCD Axions}

\author{Hai-Jun Li} 
\email{lihaijun@itp.ac.cn}
\affiliation{Key Laboratory of Theoretical Physics, Institute of Theoretical Physics, Chinese Academy of Sciences, Beijing 100190, China}
 
\author{Yu-Feng Zhou}
\email{yfzhou@itp.ac.cn}
\affiliation{Key Laboratory of Theoretical Physics, Institute of Theoretical Physics, Chinese Academy of Sciences, Beijing 100190, China}
\affiliation{School of Physical Sciences, University of Chinese Academy of Sciences, Beijing 100049, China}
\affiliation{School of Fundamental Physics and Mathematical Sciences, Hangzhou Institute for Advanced Study, UCAS, Hangzhou 310024, China}
\affiliation{International Centre for Theoretical Physics Asia-Pacific, Beijing/Hangzhou, China}

\preprint{ITP-CAS-24-166}

\date{\today}

\begin{abstract}

We introduce a novel level crossing phenomenon in the mass mixing between the QCD axions, one canonical QCD axion and one $Z_{\mathcal N}$ axion. 
The level crossing can take place at or slightly before the QCD phase transition critical temperature, depending on the ratio of the axion decay constants $\sim1.69$. 
The cosmological evolution of the mass eigenvalues in these two scenarios is similar; however, the transition of axion energy density differs significantly. 
Finally, we estimate the relic density of the QCD axion dark matter in this context. 
Additionally, this level crossing may have some interesting cosmological implications.


\end{abstract}
\maketitle


\medskip\noindent{\bf Introduction.}---%
The QCD axions are attractive candidates for the cold dark matter (DM).
The canonical QCD axion was predicted by the Peccei-Quinn (PQ) mechanism \cite{Peccei:1977hh, Peccei:1977ur} to solve the strong CP problem in the Standard Model (SM) \cite{Weinberg:1977ma, Wilczek:1977pj, Kim:1979if, Shifman:1979if, Dine:1981rt, Zhitnitsky:1980tq}.
It obtains a tiny mass from the QCD non-perturbative effects \cite{tHooft:1976rip, tHooft:1976snw}.
As the DM candidate, it can be non-thermally produced in the early Universe through the misalignment mechanism \cite{Preskill:1982cy, Abbott:1982af, Dine:1982ah}.
At high cosmic temperatures, the QCD axion was massless and obtained a non-zero mass during the QCD phase transition.
The axion oscillation when its mass is comparable to the Hubble parameter explains the observed DM abundance.
See $\rm e.g.$ Ref.~\cite{OHare:2024nmr} for a recent review.
The reduced-mass $Z_{\mathcal N}$ axion \cite{Hook:2018jle}, can also solve the strong CP problem with $\mathcal N\geqslant3$ \cite{DiLuzio:2021pxd} and account for the DM through the trapped plus kinetic misalignment mechanism \cite{DiLuzio:2021gos, Co:2019jts}.
Here $\mathcal N$ is an odd number, the $\mathcal N$ mirror worlds that are nonlinearly realized by the axion field under a $Z_{\mathcal N}$ symmetry can coexist in Nature.
Due to the suppressed non-perturbative effects on axion potential from the $\mathcal N$ degenerate QCD groups, the $Z_{\mathcal N}$ axion mass is exponentially suppressed at the QCD phase transition.
To avoid confusion, the following ``QCD axion" only stands for the canonical QCD axion.

The mass mixing in the multiple axions model \cite{Hill:1988bu} has attracted extensive attention in recent years.\footnote{For instance, the mass mixing between the QCD axion and axionlike particle (ALP) \cite{Daido:2015cba} or sterile axion \cite{Cyncynates:2023esj}, as well as the mixing between the $Z_{\mathcal N}$ axion and ALP \cite{Li:2023uvt}.}
Considering the non-zero mass mixing between these axions, the cosmological evolution process of the mass eigenvalues called the level crossing can take place and induce the adiabatic transition of the axion energy density, which is similar to the MSW effect \cite{Wolfenstein:1977ue, Mikheyev:1985zog, Mikheev:1986wj} in the neutrino oscillations.
The level crossing has some interesting cosmological implications, such as the modification of the axion relic density and isocurvature perturbations \cite{Kitajima:2014xla, Ho:2018qur, Cyncynates:2021xzw, Li:2023xkn}, the domain walls formation \cite{Daido:2015bva}, the gravitational waves emission and primordial black holes formation \cite{Cyncynates:2022wlq, Chen:2021wcf, Li:2024psa}, and also the dark energy composition \cite{Muursepp:2024mbb, Muursepp:2024kcg}.

In this letter, we investigate the mass mixing between the QCD axions, one QCD axion and one $Z_{\mathcal N}$ axion.
We find that the level crossing can take place at the QCD phase transition critical temperature or slightly before it, depending on the relationship between the axion decay constants.
The conditions for level crossing to occur in these two cases are discussed in detail.
The cosmological evolution of the mass eigenvalues in both cases is similar, whereas the transition of energy density among the axions differs.
Finally, we estimate the relic density of the QCD axion and $Z_{\mathcal N}$ axion DM through the misalignment mechanism.
To our knowledge, this is the first investigation of the effect of level crossing in the mass mixing within a multiple QCD axions model.

\medskip\noindent{\bf The model.}---%
Here we consider a minimal multiple QCD axions model, one QCD axion $\phi$ and one $Z_{\mathcal N}$ axion $\varphi$.
Considering that these two axions are charged under two copies of the SM, yet both are coupled to QCD.
We expect $\phi$ to receive the dominant contribution from the QCD potential, leading to the cancellation of the potential that occurs in the usual $Z_{\mathcal N}$ axion case, while such cancellation would not occur for the single $\varphi$.\footnote{There is a widely held belief that string theory can predict a substantial number of axions \cite{Witten:1984dg, Green:1984sg, Svrcek:2006yi}, encompassing various types such as the QCD axion and potentially the $Z_{\mathcal{N}}$ axion, known as the axiverse \cite{Arvanitaki:2009fg, Cicoli:2012sz, Demirtas:2021gsq}. In this theoretical context, a theory that incorporates the QCD axions within our framework emerges as a particularly intriguing physics scenario beyond the Standard Model (BSM). This pursuit not only enhances our understanding of their characteristics but may also uncover new physics within the BSM framework.}
The low-energy effective Lagrangian is given by
\begin{eqnarray}
\begin{aligned}
\mathcal{L}&\supset\dfrac{1}{2}\sum_{\phi, \varphi}\partial_\mu\Phi\partial^\mu\Phi-m_a^2(T) f_a^2 \left[1-\cos\left(\dfrac{\phi}{f_a}\right)\right]\\
&-\dfrac{m_{\mathcal N}^2(T) F_a^2}{{\mathcal N}^2}\left[1-\cos\left(\dfrac{\phi}{f_a}+{\mathcal N}\dfrac{\varphi}{F_a}\right)\right]\, ,
\label{eq1_mix}
\end{aligned}
\end{eqnarray}
where $m_a(T)$ and $m_{\mathcal N}(T)$ are the temperature-dependent QCD axion and $Z_{\mathcal N}$ axion masses, respectively, $f_a$ and $F_a$ are the QCD axion and $Z_{\mathcal N}$ axion decay constants.
The QCD axion mass $m_a(T)$ can be described by
\begin{eqnarray}
\begin{cases}
\dfrac{m_\pi f_\pi}{f_a}\dfrac{\sqrt{z}}{1+z}\, , & T\leq T_{\rm QCD}\\ 
\dfrac{m_\pi f_\pi}{f_a}\dfrac{\sqrt{z}}{1+z}\left(\dfrac{T}{T_{\rm QCD}}\right)^{-b}\, , & T> T_{\rm QCD} 
\end{cases} 
\label{maT}
\end{eqnarray} 
where $m_\pi$ and $f_\pi$ represent the mass and decay constant of the pion, respectively, $z\equiv m_u/m_d\simeq0.48$ signifies the ratio of the up to down quark masses, $T_{\rm QCD}\simeq150\, \rm MeV$ is the QCD phase transition critical temperature, and $b\simeq4.08$ is an index taken from the dilute instanton gas approximation.
The first term in Eq.~(\ref{maT}) corresponds to the zero-temperature QCD axion mass $m_{a,0}$.
While the $Z_{\mathcal N}$ axion mass $m_{\mathcal N}(T)$ can be described by
\begin{eqnarray}
\begin{cases}
\dfrac{m_\pi f_\pi}{\sqrt[4]{\pi} F_a}\sqrt[4]{\dfrac{1-z}{1+z}}{\mathcal N}^{3/4}z^{\mathcal N/2}\, , &T\leq T_{\rm QCD}\\
\dfrac{m_\pi f_\pi}{F_a}\sqrt{\dfrac{z}{1-z^2}}\, , &T_{\rm QCD} < T \leq \dfrac{T_{\rm QCD}}{\gamma}\\
\dfrac{m_\pi f_\pi}{F_a}\sqrt{\dfrac{z}{1-z^2}}\left(\dfrac{\gamma T}{T_{\rm QCD}}\right)^{-b}\, , &T>\dfrac{T_{\rm QCD}}{\gamma}
\end{cases} 
\label{mNT}
\end{eqnarray}
where $\gamma\in(0,1)$ is a temperature parameter.
The first term in Eq.~(\ref{mNT}) also corresponds to the zero-temperature $Z_{\mathcal N}$ axion mass $m_{{\mathcal N},0}$.
Notably, this mass undergoes a sudden exponential suppression at $T_{\rm QCD}$ due to the $Z_{\mathcal N}$ symmetry.
Subsequently, the equations of motion of $\phi$ and $\varphi$ are given by
\begin{eqnarray}
\begin{aligned}
\ddot\phi&+3H\dot\phi+m_a^2(T) f_a\sin{\left(\dfrac{\phi}{f_a}\right)}\\
&+\dfrac{m_{\mathcal N}^2(T) F_a^2}{{\mathcal N}^2 f_a} \sin\left(\dfrac{\phi}{f_a}+{\mathcal N}\dfrac{\varphi}{F_a}\right)=0\, ,
\end{aligned}
\end{eqnarray}
and
\begin{eqnarray}
\ddot\varphi+3H\dot\varphi+\dfrac{m_{\mathcal N}^2(T) F_a}{\mathcal N}\sin\left(\dfrac{\phi}{f_a}+{\mathcal N}\dfrac{\varphi}{F_a}\right)=0 \, , 
\end{eqnarray}
where $H(T)$ is the Hubble parameter, and the dots represent derivatives with respect to the physical time $t$.
Diagonalizing the mass mixing matrix
\begin{eqnarray}
\mathbf{M}^2=
\left(
\begin{array}{cc}
m_a^2(T)+\dfrac{m_{\mathcal N}^2(T) F_a^2}{{\mathcal N}^2 f_a^2}  & \quad \dfrac{m_{\mathcal N}^2(T) F_a}{{\mathcal N}f_a} \\
\dfrac{m_{\mathcal N}^2(T) F_a}{{\mathcal N}f_a} & \quad m_{\mathcal N}^2(T)
\end{array}
\right)\, ,
\end{eqnarray}
we can derive the heavy ($a_h$) and light ($a_l$) axion mass eigenstates
\begin{eqnarray}
\left(
\begin{array}{c}
a_h \\
a_l 
\end{array}
\right)
=
\left(
\begin{array}{cc}
\cos \alpha & \quad \sin \alpha \\
-\sin \alpha  & \quad   \cos \alpha
\end{array}
\right)
\left(
\begin{array}{c}
\phi \\
\varphi 
\end{array}
\right)\, ,
\end{eqnarray} 
with the corresponding mass eigenvalues $m_{h,l}(T)$
\begin{eqnarray}
\begin{aligned}
m_{h,l}^2(T)&=\dfrac{1}{2}\left[m_a^2(T)+m_{\mathcal N}^2(T)+ \dfrac{m_{\mathcal N}^2(T) F_a^2}{{\mathcal N}^2 f_a^2}\right]\\
&\pm\dfrac{1}{2{\mathcal N}^2 f_a^2}\bigg[-4 {\mathcal N}^4 m_a^2(T) m_{\mathcal N}^2(T) f_a^4\\
&+\bigg(\left(m_a^2(T)+m_{\mathcal N}^2(T)\right){\mathcal N}^2 f_a^2\\
&+m_{\mathcal N}^2(T) F_a^2\bigg)^2\bigg]^{1/2}\, ,
\end{aligned}
\end{eqnarray} 
where $\alpha$ is the mass mixing angle
\begin{eqnarray}
\cos^2\alpha=\dfrac{1}{2}\left(1+\dfrac{m_{\mathcal N}^2(T)-m_a^2(T)-\dfrac{m_{\mathcal N}^2(T) F_a^2}{{\mathcal N}^2 f_a^2}}{m_h^2(T)-m_l^2(T)}\right)\, .
\label{mixing_angle}
\end{eqnarray} 

The so-called axion level crossing is that when considering the interaction between two axion fields, there exists a nontrivial mass mixing of their temperature-dependent mass eigenvalues $m_{h,l}(T)$ that as functions of the cosmic temperature $T$.
It can take place when the difference of $m_h^2(T)-m_l^2(T)$ gets a minimum value.

\begin{figure*}[t]
\centering
\includegraphics[width=0.475\textwidth]{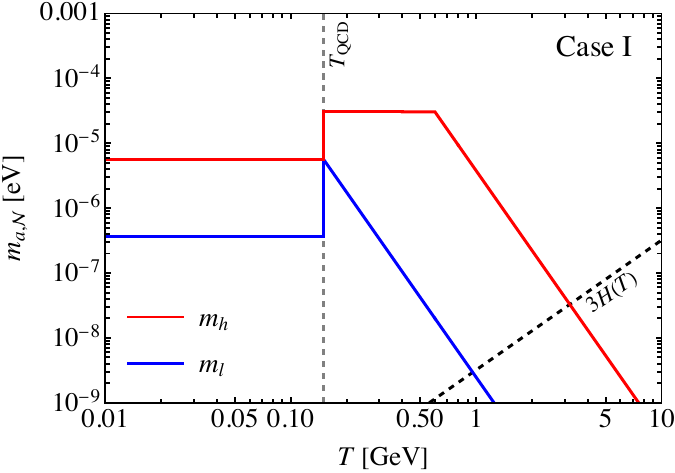}\quad\includegraphics[width=0.475\textwidth]{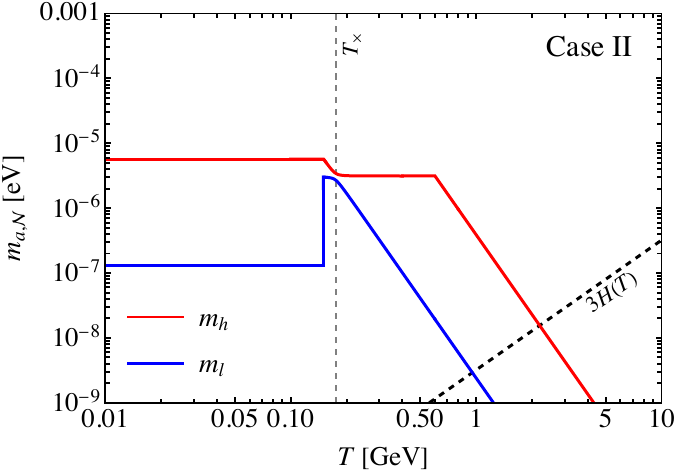}  
\caption{The illustration of level crossing in the Case~I (left) and Case~II (right).
The red and blue lines represent the mass eigenvalues $m_h(T)$ and $m_l(T)$, respectively.
The black dashed line represents the Hubble parameter $H(T)$.
Left: We set $f_a=10^{12}\,{\rm GeV}$, $F_a=10^{11.5}\,{\rm GeV}$ ($\Rightarrow\zeta\simeq 0.32$), $\mathcal N=17$, and $\gamma=0.25$.
The gray dashed line represents the temperature $T_{\rm QCD}$.
Right: We set $f_a=10^{12}\,{\rm GeV}$, $F_a=10^{12.5}\,{\rm GeV}$ ($\Rightarrow\zeta\simeq 3.16$), $\mathcal N=13$, and $\gamma=0.25$.
The gray dashed line represents the level crossing temperature $T_\times\simeq 176.3\, \rm MeV$.}
\label{fig_case}
\end{figure*}

\medskip\noindent{\bf Case~I: Level crossing at $T_{\rm QCD}$.}---%
We first consider a case that the zero-temperature QCD axion mass $m_{a,0}$ is smaller than the second term in Eq.~(\ref{mNT}) (we can define it as the mass $m_{{\mathcal N},\pi}$)
\begin{eqnarray}
\dfrac{m_\pi f_\pi}{f_a}\dfrac{\sqrt{z}}{1+z}<\dfrac{m_\pi f_\pi}{F_a}\sqrt{\dfrac{z}{1-z^2}}\, ,
\end{eqnarray}
it can be characterized as
\begin{eqnarray}
\zeta\equiv\dfrac{F_a}{f_a}<\sqrt{\dfrac{1+z}{1-z}}\simeq 1.69 \, ,
\label{con_I_1}
\end{eqnarray}
where we have defined the ratio $\zeta$.
On the other hand, in order for the level crossing to occur in this case, the zero-temperature QCD axion mass should be larger than the zero-temperature $Z_{\mathcal N}$ axion mass $m_{{\mathcal N},0}$, we have
\begin{eqnarray}
\dfrac{m_\pi f_\pi}{f_a}\dfrac{\sqrt{z}}{1+z}>\dfrac{m_\pi f_\pi}{\sqrt[4]{\pi} F_a}\sqrt[4]{\dfrac{1-z}{1+z}}{\mathcal N}^{3/4}z^{\mathcal N/2}\, ,
\end{eqnarray}
it can also be characterized as
\begin{eqnarray}
\begin{aligned}
\zeta&>\dfrac{1}{\sqrt[4]{\pi}}\sqrt[4]{\dfrac{\left(1-z\right)\left(1+z\right)^3}{z^2}}{\mathcal N}^{3/4}z^{\mathcal N/2}\\
&\simeq 1.24\times0.48^{\mathcal N/2} {\mathcal N}^{3/4}\, ,
\label{con_I_2}
\end{aligned}
\end{eqnarray}
which depends on the value of $\mathcal N$.
Then we show in Fig.~\ref{fig_case} (left) an illustration of level crossing in this case.
We have set $f_a=10^{12}\,{\rm GeV}$, $F_a=10^{11.5}\,{\rm GeV}$, $\mathcal N=17$, and $\gamma=0.25$.
The red and blue lines correspond to the temperature-dependent mass eigenvalues $m_h(T)$ and $m_l(T)$, respectively.
Note that the level crossing occurs at the QCD phase transition critical temperature
\begin{eqnarray}
T_\times = T_{\rm QCD}\, .
\end{eqnarray} 
Here we discuss the cosmological evolution of the mass eigenvalues in this scenario.
At high temperatures, the heavy mass eigenvalue $m_h(T)$ corresponds to the $Z_{\mathcal N}$ axion, whereas the light one $m_l(T)$ corresponds to the QCD axion.
They will approach to each other at $T_{\rm QCD}$ and then move away from each other.
When $T < T_{\rm QCD}$, the $m_h(T)$ represents the QCD axion, and the $m_l(T)$ represents the $Z_{\mathcal N}$ axion.
For a more intuitive description, we can express them as follows
\begin{eqnarray}
\begin{aligned}
m_h(T)\Rightarrow
\begin{cases}
m_a(T)\, , &T< T_\times \\
m_{\mathcal N}(T)\, , &T> T_\times
\end{cases} 
\end{aligned}
\end{eqnarray}
and
\begin{eqnarray}
\begin{aligned}
m_l(T)\Rightarrow
\begin{cases}
m_{\mathcal N}(T)\, , &T< T_\times \\
m_a(T)\, . &T> T_\times
\end{cases} 
\end{aligned}
\end{eqnarray}
The above discussion is focused on the Case~I, where the level crossing occurs at $T_{\rm QCD}$.

To compare with another case discussed below, we show the condition for level crossing to occur in the Case~I by using Eqs.~(\ref{con_I_1}) and (\ref{con_I_2}).
See Fig.~\ref{fig_condition} with the red shaded region in the $\{{\mathcal N}, \, \zeta\}$ plane.
The unshaded regions represent areas where no level crossing occurs.

\medskip\noindent{\bf Case~II: Level crossing slightly before $T_{\rm QCD}$.}---%
Next we consider another case that the zero-temperature QCD axion mass is larger than $m_{{\mathcal N},\pi}$, we have
\begin{eqnarray}
\zeta>\sqrt{\dfrac{1+z}{1-z}}\simeq 1.69 \, .
\label{con_II_1}
\end{eqnarray}
We present an illustration of this case in Fig.~\ref{fig_case} (right).
We have set $f_a=10^{12}\,{\rm GeV}$, $F_a=10^{12.5}\,{\rm GeV}$, $\mathcal N=13$, and $\gamma=0.25$.
Different from the Case~I, here the level crossing will take place slightly before the temperature $T_{\rm QCD}$.
By solving $d(m_h^2(T)-m_l^2(T))/dT=0$ at $T_{\rm QCD} < T < T_{\rm QCD}/\gamma$, we obtain the level crossing temperature
\begin{eqnarray}
T_\times=T_{\rm QCD} \left(\sqrt{\dfrac{1-z}{1+z}} \zeta\right)^{1/b}\left(1-\dfrac{\zeta^2}{{\mathcal N}^2}\right)^{-1/(2b)}\, ,
\label{Tx}
\end{eqnarray}
which is slightly higher than $T_{\rm QCD}$.
The QCD axion mass at $T_\times$ is given by
\begin{eqnarray}
m_a(T_\times)\simeq m_{{\mathcal N},\pi} \sqrt{1-\dfrac{\zeta^2}{{\mathcal N}^2}}\, .
\end{eqnarray}
Note that the condition $\zeta<\mathcal N$ should be satisfied in Eq.~(\ref{Tx}).
This level crossing persists for a parametric duration given by
\begin{eqnarray}
\Delta T_\times=\bigg|\dfrac{1}{\cos\alpha(T_\times)}\dfrac{d\cos\alpha(T_\times)}{dT}\bigg|^{-1}\, ,
\label{Delta_T}
\end{eqnarray} 
where $\alpha(T)$ represents the mass mixing angle given by Eq.~(\ref{mixing_angle}). 
This scenario is referred to as the Case~II.
The temperature-dependent behavior of the mass eigenvalues is similar to that in the Case~I.
However, the transition of energy density between them differs significantly, as will be discussed in detail later.
It is worth noting that the Case~I can transition into the Case~II, depending on the value of the ratio $\zeta\sim1.69$.

\begin{figure}[t]
\centering
\includegraphics[width=0.475\textwidth]{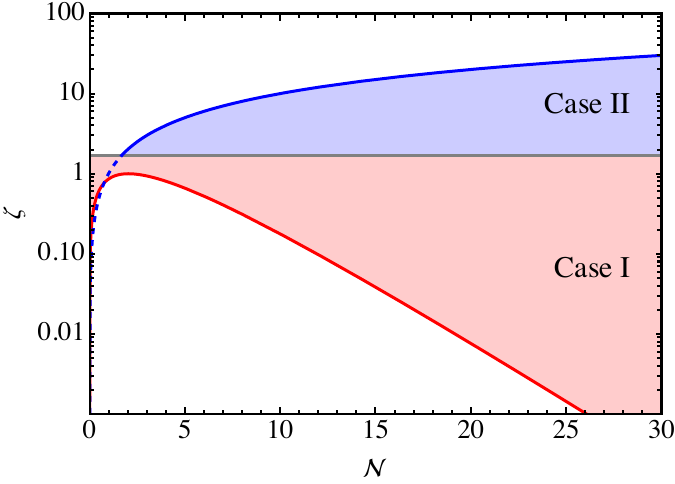}  
\caption{The conditions for level crossing to occur in the $\{{\mathcal N}, \, \zeta\}$ plane.
The red and blue shaded regions correspond to the Cases~I and II, respectively.
The gray line represents $\zeta=1.69$.
Note that ${\mathcal N}=2k+1$, $k\in\mathbb{N}^+$.}
\label{fig_condition}
\end{figure}

Then, in Fig.~\ref{fig_condition} (the blue shaded region), we illustrate the condition for level crossing to occur in the Case~II, using Eq.~(\ref{con_II_1}) and the inequality $\zeta<\mathcal{N}$.
Additionally, we note that at $T_{\rm QCD}/\gamma$ the QCD axion mass should be smaller than the $Z_{\mathcal N}$ axion mass   
\begin{eqnarray}
\dfrac{m_\pi f_\pi}{f_a}\dfrac{\sqrt{z}}{1+z}\gamma^b<\dfrac{m_\pi f_\pi}{F_a}\sqrt{\dfrac{z}{1-z^2}}\, ,
\end{eqnarray}
it can be characterized as
\begin{eqnarray}
\zeta<\sqrt{\dfrac{1+z}{1-z}}\gamma^{-b}\simeq 1.69 \, \gamma^{-b}\, ,
\label{con_II_2}
\end{eqnarray}
which depends on the value of $\gamma$.
Since for the small $\gamma$, the constraint of $\zeta<\mathcal N$ is more stringent than Eq.~(\ref{con_II_2}), we do not include it in Fig.~\ref{fig_condition}.

\medskip\noindent{\bf Axion relic density.}---%
Here we estimate the relic density of the axion DM in the Cases~I and II through the misalignment mechanism.
We will consider the pre-inflationary scenario, in which the PQ symmetry is spontaneously broken during inflation. 

\begin{figure*}[t]
\centering
\subfigcapskip=0pt
\subfigbottomskip=0pt
\subfigure[~$Z_{\mathcal N}$ axion in the Case~I.]{\includegraphics[width=5.96cm]{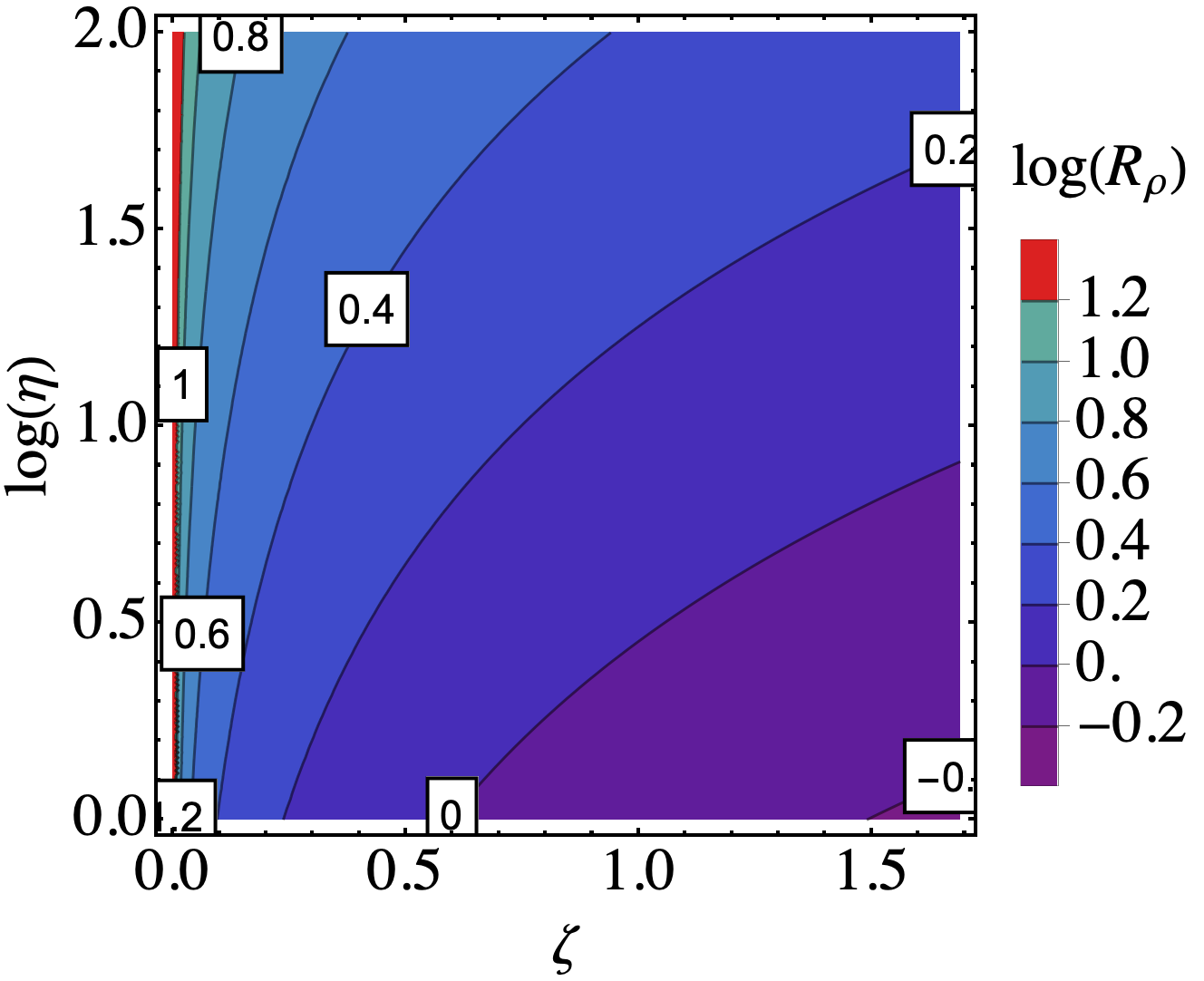}}\subfigure[~$Z_{\mathcal N}$ axion in the Case~II.]{\includegraphics[width=5.96cm]{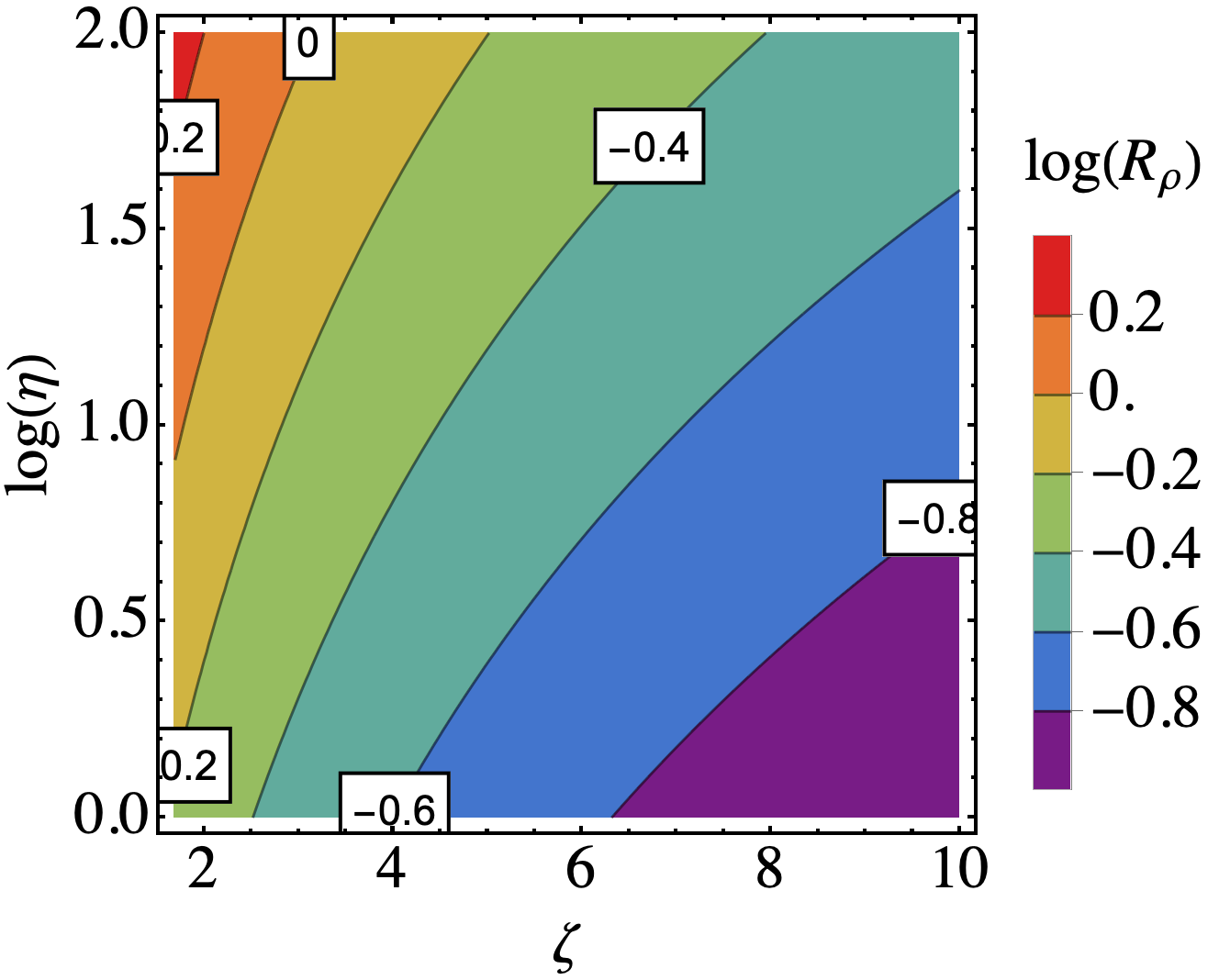}}\subfigure[~QCD axion in the Case~II.]{\includegraphics[width=5.96cm]{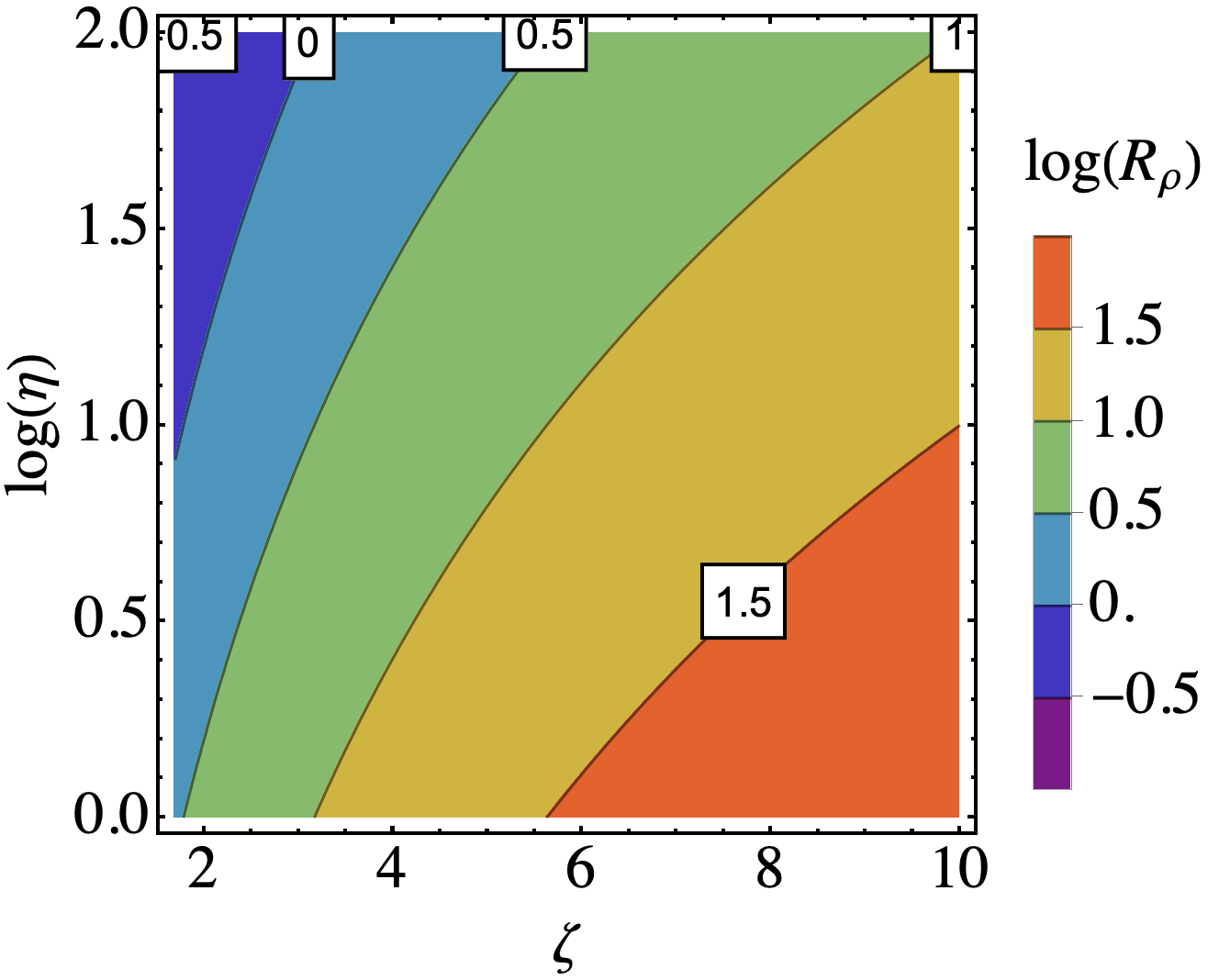}}
\caption{The distributions of the axion relic density ratio $\log(R_{\rho})$ in the $\{\zeta, \log(\eta)\}$ plane.
The left, middle, and right panels represent the $Z_{\mathcal N}$ axion in the Case~I, the $Z_{\mathcal N}$ axion in the Case~II, and the QCD axion in the Case~II, respectively.
Note that the initial misalignment angles are assumed to be of order one.}
\label{fig_relic_density}
\end{figure*}

We first discuss the Case~I, focusing on the the $Z_{\mathcal N}$ axion DM.\footnote{Since in the Case~I the QCD axion energy density at $T_{\rm QCD}$ is non-adiabatic, its relic density is not straightforward to determine. As a result, we focus our attention on the $Z_{\mathcal{N}}$ axion.}
We begin with the QCD axion field, which is frozen at high temperatures with an arbitrary initial misalignment angle $\theta_{1,a}$, and starts to oscillate at $T_{1,a}$.
The oscillation temperature $T_1$ is given by $m(T)=3H(T)$.
In the following, the subscript ``$x$" stands for the physical quantity at $T_x$ or corresponding to $x$.
The initial energy density in the QCD axion field is given by
\begin{eqnarray}
\rho_{a,1}=\dfrac{1}{2}m_{a,1}^2 f_a^2 \theta_{1,a}^2\, .
\label{initial_QCD}
\end{eqnarray}
At $T_{\rm QCD}<T<T_{1,a}$, the QCD axion energy density is adiabatic invariant with the comoving number $N_a \equiv \rho_a a^3 /m_a$, where $a$ is the scale factor.
Then we have the QCD axion energy density just before $T_{\rm QCD}$ as
\begin{eqnarray}
\rho_{a, \rm QCD}=\dfrac{1}{2} m_{a,0} m_{a,1} f_a^2 \theta_{1,a}^2 \left(\dfrac{a_{1,a}}{a_{\rm QCD}}\right)^3 \, .
\end{eqnarray}
In this case, the axion is trapped around $\theta_{\rm tr}\equiv\theta_a(T_{\rm QCD})$ until $T_{\rm QCD}$ with the initial axion velocity $\dot{\theta}_{\rm tr}\equiv\dot{\theta}_a(T_{\rm QCD})$.
At $T=T_{\rm QCD}$, the light mass eigenvalue $m_l(T)$ will comprise the $Z_{\mathcal N}$ axion, the axion mass is suddenly suppressed and the true minimum develops.
Then we have the mean velocity of the $Z_{\mathcal N}$ axion
\begin{eqnarray}
\sqrt{\langle\dot{\theta}_{\rm tr}^2\rangle}=\dfrac{1}{\sqrt {2}\zeta} \sqrt{m_{a,0} m_{a,1}} \theta_{1,a} \left(\dfrac{a_{1,a}}{a_{\rm QCD}}\right)^{3/2}\, .
\label{mean_velocity_ZN}
\end{eqnarray}
Note that the $Z_{\mathcal N}$ axion energy density at $T_{\rm QCD}$ is non-adiabatic, which just after $T_{\rm QCD}$ is given by
\begin{eqnarray}
\rho_{{\mathcal N}, \rm tr}=\dfrac{1}{2} F_a^2 \dot{\theta}_{\rm tr}^2+2\dfrac{m_{\mathcal N}^2 F_a^2}{{\mathcal N}^2}\, ,
\end{eqnarray}
where we consider a case that the axion has a large mean velocity.
Then the $Z_{\mathcal N}$ axion will start to oscillate at $T_2$, which is defined as the temperature when the kinetic energy is equal to the barrier height.
At $T_2<T<T_{\rm QCD}$, the $Z_{\mathcal N}$ axion energy density is conserved with the comoving PQ charge $q_{\rm kin}=\dot{\theta}_{\mathcal N} a^3$, and we have the scale factor at $T_2$ as $a_2=({\mathcal N}\dot{\theta}_{\rm tr}/(2m_{\mathcal N}))^{1/3}a_{\rm QCD}$.
Then at $T_0<T<T_2$, the adiabatic approximation is valid, and we have the $Z_{\mathcal N}$ axion energy density at present $T_0$ as
\begin{eqnarray}
\rho_{{\mathcal N},0} = \dfrac{m_{{\mathcal N},0} F_a^2 \dot{\theta}_{\rm tr}}{\mathcal N} \left(\dfrac{a_{\rm QCD}}{a_0}\right)^3\, .
\label{rho_ZN_0}
\end{eqnarray}
Substituting the mean velocity, it reads
\begin{eqnarray}
\rho_{{\mathcal N},0} = C \dfrac{m_{{\mathcal N},0} \sqrt{m_{a,0} m_{a,1}} \theta_{1,a} F_a^2}{\sqrt{2}{\mathcal N}\zeta} \left(\dfrac{\sqrt{a_{1,a} a_{\rm QCD}}}{a_0}\right)^3\, , ~
\end{eqnarray}
where $C\simeq2$ is a constant.
Compared with the no mixing case, the relic density can be modified by
\begin{eqnarray} 
R_{\rho}\sim \sqrt[4]{\dfrac{1-z}{1+z}} \sqrt[4]{\dfrac{m_{{\mathcal N},1}}{m_{a,1}}} \dfrac{\theta_{1,a}}{|\theta_{1,{\mathcal N}}-\pi| \sqrt{\zeta}}\, .
\label{ratio_1}
\end{eqnarray}
Then, in the left panel of Fig.~\ref{fig_relic_density}, we show the distribution of $R_{\rho}$ in the $\{\zeta, \log(\eta)\}$ plane, where $\eta \equiv m_{{\mathcal N},1}/m_{a,1}$ and the initial misalignment angles are assumed to be of order one.  
Considering the significant impact of the temperature parameter $\gamma$ on the oscillation temperature of the $Z_{\mathcal N}$ axion, we incorporate the variations in $\gamma$ into the observed changes in $\eta$ to provide a more intuitive illustration of how the relic density varies across different parametric conditions.    
We find that the ratio $R_{\rho}$ is enhanced in most regions, whereas it is suppressed in a small subset of regions.  
Notice that, in the Case~I, the parameter ${\mathcal N}$ also exhibits a significant property related to $\zeta$, as illustrated in Fig.~\ref{fig_condition}. 
However, this property cannot be directly observed from Fig.~\ref{fig_relic_density}, and here we roughly set $0<\zeta<1.69$.

Next we discuss the $Z_{\mathcal N}$ axion DM in the Case~II.
Because of the insights gained from the discussion in the last paragraph, we can simplify our discussion here somewhat.
We also begin with the QCD axion field, and its initial energy density at $T_{1,a}$ is given by Eq.~(\ref{initial_QCD}).
Then at $T_\times<T<T_{1,a}$, the QCD axion energy density is adiabatic invariant, and it at $T_\times$ can be described by 
\begin{eqnarray}
\rho_{a, \times}=\dfrac{1}{2} m_{{\mathcal N},\pi} m_{a,1} f_a^2 \theta_{1,a}^2 \left(\dfrac{a_{1,a}}{a_\times}\right)^3 \, .
\end{eqnarray}
At $T=T_\times$, the level crossing occurs, the light mass eigenvalue $m_l(T)$ will comprise the $Z_{\mathcal N}$ axion and the energy density $\rho_{a, \times}$ is transferred to the $Z_{\mathcal N}$ axion $\rho_{{\mathcal N},\times}$.
To ensure this axion energy transition is adiabatic at $T_\times$, we require the following condition to be satisfied
\begin{eqnarray}
\Delta t_\times \gg \max\left[\dfrac{2\pi}{m_l(T_\times)}, \, \dfrac{2\pi}{m_h(T_\times)-m_l(T_\times)}\right]\, ,
\label{Delta_time}
\end{eqnarray}
where $\Delta t_\times$ is the time interval corresponding to the temperature change described in Eq.~(\ref{Delta_T}).
Additionally, we have verified this condition using the parameter set illustrated in the right panel of Fig.~\ref{fig_case}, and the results are presented in Fig.~\ref{fig_deltat}.
Then at $T_{\rm QCD}<T<T_\times$, the adiabatic approximation is valid again, and the $Z_{\mathcal N}$ axion energy density just before $T_{\rm QCD}$ is given by
\begin{eqnarray}
\rho_{{\mathcal N}, \rm QCD}=\dfrac{1}{2} m_{{\mathcal N},\pi} m_{a,1} f_a^2 \theta_{1,a}^2 \left(\dfrac{a_{1,a}}{a_{\rm QCD}}\right)^3 \, ,
\end{eqnarray}
with the mean velocity
\begin{eqnarray}
\sqrt{\langle\dot{\theta}_{\rm tr}^2\rangle}=\dfrac{1}{\sqrt {2}\zeta} \sqrt{m_{{\mathcal N},\pi} m_{a,1}} \theta_{1,a} \left(\dfrac{a_{1,a}}{a_{\rm QCD}}\right)^{3/2}\, .
\label{mean_velocity_ZN}
\end{eqnarray}
The subsequent steps ($T<T_{\rm QCD}$) are similar to those discussed in the previous paragraph.
Substituting the mean velocity into Eq.~(\ref{rho_ZN_0}), we have the present $Z_{\mathcal N}$ axion energy density
\begin{eqnarray}
\rho_{{\mathcal N},0} = C \dfrac{m_{{\mathcal N},0} \sqrt{m_{{\mathcal N},\pi} m_{a,1}} \theta_{1,a} F_a^2}{\sqrt{2}{\mathcal N}\zeta} \left(\dfrac{\sqrt{a_{1,a} a_{\rm QCD}}}{a_0}\right)^3\, . ~
\end{eqnarray}
Now compared with the no mixing case, we find that the relic density can be modified by
\begin{eqnarray} 
R_{\rho}\sim \sqrt[4]{\dfrac{m_{{\mathcal N},1}}{m_{a,1}}} \dfrac{\theta_{1,a}}{ |\theta_{1,{\mathcal N}}-\pi| \zeta}\, .
\label{ratio_2}
\end{eqnarray}
As previously demonstrated, in the middle panel of Fig.~\ref{fig_relic_density}, we present the distribution of $R_{\rho}$ in the $\{\zeta, \log(\eta)\}$ plane. 
In most regions, the ratio $R_{\rho}$ can be suppressed, whereas in a small subset of regions, it can be enhanced.
Note also that $\zeta$ is constrained by the set of ${\mathcal N}$, and here we only present a small range of zeta values, specifically $1.69<\zeta<10$.
On the other hand, we note that Eq.~(\ref{ratio_2}) resembles the single level crossing scenario, where the $Z_{\mathcal{N}}$ axion mixes with ALP, leading to the suppression of the $Z_{\mathcal N}$ axion relic density, as seen in Eq.~(25) of Ref.~\cite{Li:2023uvt}. 
However, due to the differing ranges of parameters, here $\zeta$ exceeds 1.69, and consequently, the $Z_{\mathcal N}$ axion relic density ratio $R_{\rho}$ can be either enhanced or suppressed.

\begin{figure}[t]
\centering
\includegraphics[width=0.475\textwidth]{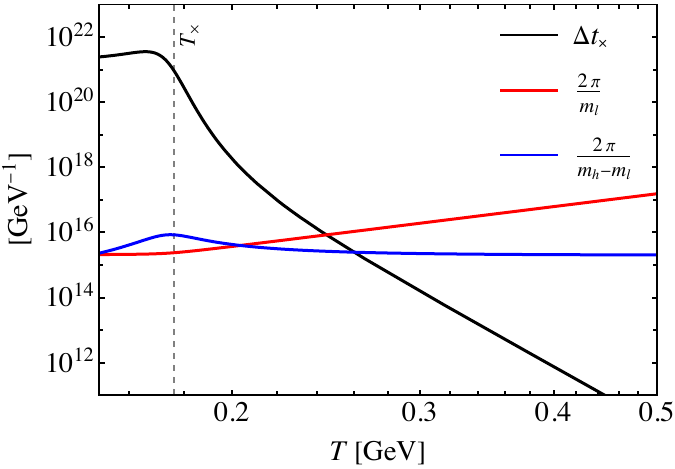}  
\caption{The three terms in Eq.~(\ref{Delta_time}) as functions of the temperature $T$.
The black, red, and blue lines represent $\Delta t_\times$, $2\pi/m_l(T)$, and $2\pi/(m_h(T)-m_l(T))$, respectively.
The gray dashed line represents $T_\times\simeq 176.3\, \rm MeV$.}
\label{fig_deltat}
\end{figure} 

Finally we discuss the QCD axion DM in the Case~II.
We should begin with the $Z_{\mathcal N}$ axion field, it starts to oscillate at $T_{1,{\mathcal N}}$ with the initial misalignment angle $\theta_{1,{\mathcal N}}$, and the initial energy density is given by
\begin{eqnarray}
\rho_{{\mathcal N},1}=\dfrac{1}{2}m_{{\mathcal N},1}^2 F_a^2 \theta_{1,{\mathcal N}}^2\, .
\end{eqnarray}
At $T_\times<T<T_{1,{\mathcal N}}$, the $Z_{\mathcal N}$ axion energy density is adiabatic invariant, and at $T_\times$ we have 
\begin{eqnarray}
\rho_{{\mathcal N},\times}=\dfrac{1}{2} m_{{\mathcal N},\pi} m_{{\mathcal N},1} F_a^2 \theta_{1,{\mathcal N}}^2 \left(\dfrac{a_{1,{\mathcal N}}}{a_\times}\right)^3 \, .
\end{eqnarray}
At $T=T_\times$, the level crossing occurs, the heavy mass eigenvalue $m_h(T)$ will comprise the QCD axion and the energy density $\rho_{{\mathcal N},\times}$ is transferred to the QCD axion $\rho_{a,\times}$.
Then at $T_0<T<T_\times$, the adiabatic approximation is valid again, and we have the present QCD axion energy density 
\begin{eqnarray}
\rho_{a,0}=\dfrac{1}{2} m_{a,0} m_{{\mathcal N},1} F_a^2 \theta_{1,{\mathcal N}}^2 \left(\dfrac{a_{1,{\mathcal N}}}{a_0}\right)^3 \, .
\end{eqnarray}
Compared with the no mixing case, the relic density can be modified by
\begin{eqnarray} 
R_{\rho}\sim \sqrt{\dfrac{m_{a,1}}{m_{{\mathcal N},1}}} \dfrac{\theta_{1,{\mathcal N}}^2 \zeta^2}{\theta_{1,a}^2}\, .
\label{ratio_3}
\end{eqnarray}
In the right panel of Fig.~\ref{fig_relic_density}, we also present the distribution of $R_{\rho}$ in the $\{\zeta, \log(\eta)\}$ plane.
We find that, within the parameter space we consider, the ratio $R_{\rho}$ is enhanced in most regions, whereas it is suppressed in a minority of those regions.
We also note that Eq.~(\ref{ratio_3}) resembles the level crossing that occurs in the mixing between the QCD axion and ALP, as illustrated in Eq.~(3.26) of Ref.~\cite{Li:2023xkn}.
In such case, the QCD axion relic density can be significantly suppressed.
However, as mentioned earlier, due to the varying ranges of parameters, the QCD axion relic density ratio $R_{\rho}$ in this context can be either enhanced or suppressed.

\medskip\noindent{\bf Conclusion.}---%
In summary, we have investigated the mass mixing between the canonical QCD axion and $Z_{\mathcal N}$ axion, and introduced a novel level crossing phenomenon.
During the mixing, we find that the level crossing can take place at the QCD phase transition critical temperature $T_{\rm QCD}$ (Case~I) or slightly before $T_{\rm QCD}$ (Case~II), depending on the ratio of the axion decay constants $\zeta\sim1.69$.
The conditions for level crossing to occur in the Cases~I and II are discussed in detail.
In the Case~I, the zero-temperature QCD axion mass $m_{a,0}$ should be smaller than the mass $m_{{\mathcal N},\pi}$ and larger than the zero-temperature $Z_{\mathcal N}$ axion mass $m_{{\mathcal N},0}$, which can be characterized by the inequality $1.24\times0.48^{\mathcal N/2} {\mathcal N}^{3/4}<\zeta<1.69$.
While in the Case~II, the zero-temperature QCD axion mass $m_{a,0}$ should be larger than $m_{{\mathcal N},\pi}$ and the condition $\zeta<\mathcal N$ should be satisfied, which can also be characterized by the inequality $1.69<\zeta<{\mathcal N}$.
Note that in this case we also have a relatively weak constraint given by $\zeta<1.69 \, \gamma^{-b}$, which will become more stringent for large values of $\gamma$.

Despite the similarity in the cosmological evolution of mass eigenvalues in both cases, the transition of energy density between the axions differs. 
Specifically, in the Case~I, the axion energy transition at the level crossing is non-adiabatic (at $T_{\rm QCD}$), while in the Case~II, it is considered to be adiabatic (at $T_\times$).
Finally, we estimate the relic density of the axion DM through the misalignment mechanism, focusing on the $Z_{\mathcal N}$ axion DM in the Cases~I and II, respectively, as well as the QCD axion DM in the Cases~II.
We also compare this with other axion mixing scenarios and find that, within the parameter space we consider, the axion relic density can be either enhanced or suppressed.
Furthermore, the level crossing phenomenon in this scenario may have other intriguing cosmological implications, such as gravitational waves and primordial black holes.

\medskip\noindent{\bf Acknowledgments.}---%
This work was supported by the CAS Project for Young Scientists in Basic Research YSBR-006, the National Key R\&D Program of China (Grant No.~2017YFA0402204), and the National Natural Science Foundation of China (Grants No.~11821505, No.~11825506, and No.~12047503).

\bibliography{references}

\clearpage
\maketitle
\onecolumngrid
\begin{center}
\textbf{\large Mass Mixing between QCD Axions}\\ 
\vspace{0.05in}
{ \it \large Supplemental Material}\\ 
\vspace{0.05in}
{Hai-Jun Li and Yu-Feng Zhou}
\end{center}
\onecolumngrid
\setcounter{equation}{0}
\setcounter{figure}{0}
\setcounter{table}{0}
\setcounter{section}{0}
\makeatletter
\renewcommand{\theequation}{S\arabic{equation}}
\renewcommand{\thefigure}{S\arabic{figure}}

This Supplemental Material is organized as follows. 
In Sec.~\ref{sm_sec1}, we provide an illustrative example of the UV completion for the $Z_{\mathcal N}$ axion.
In Sec.~\ref{sm_sec2}, we present the temperature-dependent mass of the $Z_{\mathcal N}$ axion.
In Sec.~\ref{sm_sec3}, we describe how our axion mixing potential is derived.
Finally, in Sec.~\ref{sm_sec4} we present some details about the axion relic density, considering both the no-mixing and mixing scenarios.

\section{UV completion for the $Z_{\mathcal N}$ axion}
\label{sm_sec1}

Here we provide an illustrative example of the UV completion for the $Z_{\mathcal N}$ axion.
We consider the KSVZ $Z_{\mathcal N}$ axion scenario \cite{DiLuzio:2021pxd}, with ${\mathcal N}$ copies of vector-like Dirac fermions $Q_k$ and a gauge singlet complex scalar $S$, the $Z_{\mathcal N}$ symmetry reads $Q_k\to Q_{k+1}$, $S\to e^{2\pi i/{\mathcal N}}S$.
The general Lagrangian is given by 
\begin{eqnarray}
\mathcal{L}_{\rm UV}=\partial_\mu S \partial^\mu S+\sum_{k=0}^{{\mathcal N}-1}\left[\bar{Q}_k i \slashed{\mathcal{D}}Q_k+y e^{2\pi k i/{\mathcal N}} S \bar{Q}_k \dfrac{1+\gamma_5}{2} Q_k +{\rm h.c.}\right]-V(S, H_k)\, ,
\end{eqnarray}
which exhibits an accidental U(1)$_{\rm PQ}$ symmetry that is spontaneously broken by the vacuum expectation value $v_S$ through the potential $V(S, H_k)$.
The complex scalar $S$ can be expressed as
\begin{eqnarray}
S=\dfrac{1}{\sqrt{2}}\left(v_S+\rho\right)e^{i \frac{\varphi}{v_S}}\, ,
\end{eqnarray}
where $\rho$ and $\varphi$ represent the radial and angular (axion) modes, respectively.
Through an axial transformation, the axion mode can be rotated away from the Yukawa term.
After integrating out the heavy quarks, the resulting effective Lagrangian for the axion is given by
\begin{eqnarray}
\delta\mathcal{L}_{\rm UV}=\sum_{k=0}^{{\mathcal N}-1}\dfrac{\alpha_s}{8\pi}\left(\dfrac{\varphi}{F_a}+\dfrac{2\pi k}{{\mathcal N}}\right) G_k \widetilde{G}_k\, ,
\end{eqnarray}
where we set $v_S=F_a$, $F_a$ is the axion decay constant, $\alpha_s$ is the strong fine structure constant, $G_k$ and $\widetilde{G}_k$ are the gluon field strength tensor and dual tensor, respectively. 

In the context of the KSVZ $Z_{\mathcal N}$ axion model, the discrete symmetry is inherently
present due to its construction. 
By elevating the inherent $Z_{\mathcal N}$ symmetry to a gauge symmetry, an accidental U(1)$_{\rm PQ}$ invariance arises, which, for large values of ${\mathcal N}$, is efficiently shielded from additional sources of explicit breaking. 
Furthermore, the lowest-dimensional PQ-violating operator in the scalar potential that is compatible with the $Z_{\mathcal N}$ symmetry is $S^{\mathcal N}$, which contributes to an explicitly PQ-breaking term to the potential.

See also Ref.~\cite{DiLuzio:2021pxd} for another UV completion involving the composite $Z_{\mathcal N}$ axion model.

\section{Temperature-dependent $Z_{\mathcal N}$ axion mass}
\label{sm_sec2}
Here we provide a brief overview of the temperature-dependent $Z_{\mathcal N}$ axion mass.
As derived in Ref.~\cite{DiLuzio:2021pxd}, the zero-temperature ($T\leq T_{\rm QCD}$) $Z_{\mathcal N}$ axion potential is given by 
\begin{eqnarray}
V_{\mathcal N}(\theta)\simeq -\dfrac{m_{{\mathcal N},0}^2 F_a^2}{{\mathcal N}^2} \cos\left({\mathcal N}\theta\right)\, ,
\end{eqnarray}
with the zero-temperature axion mass
\begin{eqnarray}
m_{{\mathcal N},0}\simeq \dfrac{m_\pi f_\pi}{\sqrt[4]{\pi} F_a}\sqrt[4]{\dfrac{1-z}{1+z}}{\mathcal N}^{3/4}z^{\mathcal N/2}\, ,
\end{eqnarray}
where $z\equiv m_u/m_d\simeq0.48$.
At medium temperatures $T_{\rm QCD}<T\leq T_{\rm QCD}/\gamma$, the $Z_{\mathcal N}$ axion potential is given by
\begin{eqnarray}
V_{\mathcal N}(\theta)\simeq V(\theta) \left(\dfrac{T}{T_{\rm QCD}}\right)^{-2b} + \sum_{k=1}^{{\mathcal N}-1}V\left(\theta+\dfrac{2\pi k}{{\mathcal N}}\right)\, ,
\end{eqnarray}
where $V(\theta)$ is the QCD axion zero-temperature chiral potential
\begin{eqnarray}
V(\theta)=-m_\pi^2 f_\pi^2 \sqrt{1-\dfrac{4z}{\left(1+z\right)^2}\sin^2\left(\dfrac{\theta}{2}\right)} \, .
\end{eqnarray}
Using 
\begin{eqnarray}
m_{\mathcal N}(T)=\dfrac{1}{F_a}\sqrt{\dfrac{d^2 V_{\mathcal N}(\theta)}{d \theta^2}\bigg|_{\rm min}}\, ,
\end{eqnarray}
we obtain the axion mass
\begin{eqnarray}
m_{\mathcal N}\simeq \dfrac{m_\pi f_\pi}{F_a}\sqrt{\dfrac{z}{1-z^2}}\, ,
\end{eqnarray}
which is also defined as the mass $m_{{\mathcal N},\pi}$.
At high temperatures $T>T_{\rm QCD}/\gamma$, the $Z_{\mathcal N}$ axion potential is given by
\begin{eqnarray}
V_{\mathcal N}(\theta)\simeq V(\theta) \left(\dfrac{T}{T_{\rm QCD}}\right)^{-2b} + \sum_{k=1}^{{\mathcal N}-1}V\left(\theta+\dfrac{2\pi k}{{\mathcal N}}\right) \left(\dfrac{\gamma T}{T_{\rm QCD}}\right)^{-2b}\, ,
\end{eqnarray}
then we have the axion mass 
\begin{eqnarray}
m_{\mathcal N}\simeq \dfrac{m_\pi f_\pi}{F_a}\sqrt{\dfrac{z}{1-z^2}}\left(\dfrac{\gamma T}{T_{\rm QCD}}\right)^{-b}\, .
\end{eqnarray}
Notice that the temperature parameter $\gamma$ delineates the high-temperature and intermediate-temperature regions.
To be more precise, this parameter can be represented by the inequality $\gamma_{\rm min} < \gamma < \gamma_{\rm max}$, where the values of $\gamma$ at the lower and upper bounds respectively correspond to the mirror copies of the SM that are the coldest and the hottest. 
For simplicity, we denote it as $\gamma$.
Finally, the temperature-dependent $Z_{\mathcal N}$ axion mass can be described by
\begin{eqnarray}
m_{\mathcal N}(T)\simeq
\begin{cases}
\dfrac{m_\pi f_\pi}{\sqrt[4]{\pi} F_a}\sqrt[4]{\dfrac{1-z}{1+z}}{\mathcal N}^{3/4}z^{\mathcal N/2}\, , &T\leq T_{\rm QCD}\\
\dfrac{m_\pi f_\pi}{F_a}\sqrt{\dfrac{z}{1-z^2}}\, , &T_{\rm QCD} < T \leq \dfrac{T_{\rm QCD}}{\gamma}\\
\dfrac{m_\pi f_\pi}{F_a}\sqrt{\dfrac{z}{1-z^2}}\left(\dfrac{\gamma T}{T_{\rm QCD}}\right)^{-b}\, . &T>\dfrac{T_{\rm QCD}}{\gamma}
\end{cases} 
\end{eqnarray}
See also Ref.~\cite{DiLuzio:2021gos} for more details.

\section{Axion mixing potential}
\label{sm_sec3}

In this section, we describe how our axion mixing potential, which is presented in Eq.~(\ref{eq1_mix}), is derived.
Let us first consider the mixing of two axions $\phi_1$ and $\phi_2$ with the mixing potential
\begin{eqnarray}
V_{\rm mix}=\Lambda_1^4\left[1-\cos\left(n_{11}\dfrac{\phi_1}{f_1}+n_{12}\dfrac{\phi_2}{f_2}+\delta_1\right)\right]+\Lambda_2^4\left[1-\cos\left(n_{21}\dfrac{\phi_1}{f_1}+n_{22}\dfrac{\phi_2}{f_2}+\delta_2\right)\right]\, ,
\end{eqnarray}
where $n_{ij}$ are the domain wall numbers, and $\delta_{i}$ are the constant phases.
In order to achieve the purpose of discussing the emergence of the level crossing phenomenon, the domain wall numbers must be set in this way:
\begin{eqnarray}
n_{11}\neq 0\, (n_{11}=0)\, ,\quad n_{12}=0\, (n_{12}\neq 0)\, ,\quad n_{21}\neq 0\, ,\quad n_{22}\neq0\, ,
\label{domain_wall_1}
\end{eqnarray}
or
\begin{eqnarray}
n_{11}\neq 0\, ,\quad n_{12}\neq0\, ,\quad n_{21}= 0\, (n_{21}\neq0)\, ,\quad n_{22}\neq0\, (n_{22}= 0)\, .
\label{domain_wall_2}
\end{eqnarray}
In other words, among these domain wall numbers, one must be set to zero, while the rest cannot be zero.
This consideration and operation are reasonable, and numerous pieces of literature considering the mixing between two axions have adopted similar approaches.
For instance, the mixing between the QCD axion and ALP \cite{Daido:2015cba} or sterile axion \cite{Cyncynates:2023esj}, as well as the mixing between the $Z_{\mathcal N}$ axion and ALP \cite{Li:2023uvt}.
Note that Refs.~\cite{Daido:2015cba} and \cite{Cyncynates:2023esj} can be viewed as corresponding to the considerations in the contexts of Eqs.~(\ref{domain_wall_1}) and (\ref{domain_wall_2}), respectively, if we consider $\phi_1$ as the QCD axion and $\phi_2$ as the ALP.
In these contexts, the domain wall numbers are equal to or less than 1 (and there is exactly one that is less than 1, $\rm i.e.$, 0). 
In contrast, in Ref.~\cite{Li:2023uvt} the domain wall numbers can be greater than 1.
On the other hand, the constant phases need to be set to zero
\begin{eqnarray}
\delta_1=0\, ,\quad \delta_2=0\, .
\end{eqnarray}
This can be regarded as tantamount to demanding an independent solution for addressing the strong CP problem. 
Since this assumption has no impact on the evolution of the two axion fields during the mixing process, it will not alter the energy density of axions in the context of mass mixing \cite{Li:2023xkn}.

Now consider the scenario of multiple QCD axions mixing in this work, specifically one QCD axion $\phi$ and one $Z_{\mathcal N}$ axion $\varphi$. 
For our purposes, we need to consider the following parameter settings:
\begin{eqnarray}
n_{11}=1\, ,\quad n_{12}=0\, ,\quad n_{21}=1\, ,\quad n_{22}={\mathcal N}\, ,\quad \delta_1=0\, ,\quad \delta_2=0\, ,
\end{eqnarray}
therefore we obtain the mixing potential in Eq.~(\ref{eq1_mix}), which is given by
\begin{eqnarray}
V_{\rm mix}=m_a^2(T) f_a^2 \left[1-\cos\left(\dfrac{\phi}{f_a}\right)\right]+\dfrac{m_{\mathcal N}^2(T) F_a^2}{{\mathcal N}^2}\left[1-\cos\left(\dfrac{\phi}{f_a}+{\mathcal N}\dfrac{\varphi}{F_a}\right)\right]\, .
\label{mixing_potential}
\end{eqnarray}
As mentioned above, obtaining the axion mixing potential here is both straightforward and reasonable.
When we set $n_{11}=1$ and $n_{12}=0$, the first term of Eq.~(\ref{mixing_potential}) aligns perfectly with the canonical QCD axion potential; however, this alignment is not an absolute requirement in this context.
Additionally, it is worth noting that the acquisition of the mixing potential here differs from the acquisition of the single $Z_{\mathcal N}$ axion potential.
Furthermore, there should also exist another mixing scenario with the potential given by
\begin{eqnarray}
V_{\rm mix}=m_a^2(T) f_a^2 \left[1-\cos\left(\dfrac{\phi}{f_a}+{\mathcal N}\dfrac{\varphi}{F_a}\right)\right]+\dfrac{m_{\mathcal N}^2(T) F_a^2}{{\mathcal N}^2}\left[1-\cos\left({\mathcal N}\dfrac{\varphi}{F_a}\right)\right]\, .
\end{eqnarray}
The corresponding cosmological implications would be entirely distinct and warrant a separate detailed discussion, which, however, is beyond the scope of this particular context.

\section{Axion relic density}
\label{sm_sec4}

In this section, we present some details about the axion relic density through the misalignment mechanism.
We consider the pre-inflationary scenario, in which the PQ symmetry is spontaneously broken during inflation.

\medskip\noindent{\bf QCD axion without mixing.}---%
We first show the QCD axion relic density without mixing.
At high temperatures $T\gg T_{\rm QCD}$, the QCD axion field is frozen at an arbitrary initial misalignment angle $\theta_{1,a}$ and starts to oscillate at $T_{1,a}$.
The oscillation temperature $T_{1,a}$ is given by
\begin{eqnarray}
m(T)=3H(T)\, ,
\end{eqnarray}
with the Hubble parameter
\begin{eqnarray}
H(T)=\sqrt{\dfrac{\pi^2 g_*(T)}{90}}\dfrac{T^2}{M_{\rm Pl}}\, ,
\end{eqnarray} 
the number of effective degrees of freedom of the energy density $g_*$, and the reduced Planck mass $M_{\rm Pl}\simeq 2.4\times10^{18}\, \rm GeV$.
The initial energy density in the QCD axion field is 
\begin{eqnarray}
\rho_{a,1}=\dfrac{1}{2}m_{a,1}^2 f_a^2 \theta_{1,a}^2\, ,
\end{eqnarray}
where $m_{a,1}$ is the axion mass at $T_{1,a}$.
At $T_0<T<T_{1,a}$, the QCD axion energy density is adiabatic invariant with the comoving number $N_a \equiv \rho_a a^3 /m_a$, where $a$ is the scale factor.
Using 
\begin{eqnarray}
\dfrac{\rho_{a,1} a_{1,a}^3}{m_{a,1}}=\dfrac{\rho_{a,0} a_0^3}{m_{a,0}}\, ,
\end{eqnarray}
then the QCD axion energy density at present ($T_0$) can be described by
\begin{eqnarray}
\rho_{a,0} =\dfrac{1}{2} m_{a,0} m_{a,1} f_a^2 \theta_{1,a}^2 \left(\dfrac{a_{1,a}}{a_0}\right)^3\, ,
\end{eqnarray}
where $a_{1,a}$ and $a_0$ are the scale factors at $T_{1,a}$ and $T_0$, respectively.

\medskip\noindent{\bf $Z_{\mathcal N}$ axion without mixing.}---%
Next we show the $Z_{\mathcal N}$ axion relic density without mixing.
At high temperatures $T> T_{\rm QCD}$, the $Z_{\mathcal N}$ axion field is frozen at an initial misalignment angle $\theta_{1,{\mathcal N}}$ and starts to oscillate at the oscillation temperature $T_{1,{\mathcal N}}$. 
Here we consider a case that the $Z_{\mathcal N}$ axion mass is temperature-dependent at the first oscillation, $\rm i.e.$, $T_{1,{\mathcal N}}>1/\gamma T_{\rm QCD}$.
The initial energy density in the $Z_{\mathcal N}$ axion field is 
\begin{eqnarray}
\rho_{{\mathcal N},1}=\dfrac{1}{2}m_{{\mathcal N},1}^2 F_a^2 \theta_{1,{\mathcal N}}^2\, ,
\end{eqnarray}
where $m_{{\mathcal N},1}$ is the axion mass at $T_{1,{\mathcal N}}$.
At $T_{\rm QCD}<T<T_{1,{\mathcal N}}$, the $Z_{\mathcal N}$ axion energy density is adiabatic invariant with the comoving number $N_{\mathcal N} \equiv \rho_{\mathcal N} a^3 /m_{\mathcal N}$.
Then we have the $Z_{\mathcal N}$ axion energy density at $T_{\rm QCD}$ as
\begin{eqnarray}
\rho_{{\mathcal N}, \rm QCD}=\dfrac{1}{2} m_{{\mathcal N},1} m_{{\mathcal N},\pi}  F_a^2 \left(\theta_{1,{\mathcal N}}-\pi\right)^2 \left(\dfrac{a_{1,{\mathcal N}}}{a_{\rm QCD}}\right)^3 \, ,
\end{eqnarray}
where $a_{{\mathcal N},a}$ is the scale factor at $T_{1,{\mathcal N}}$.
In this period, the $Z_{\mathcal N}$ axion will be trapped around $\theta_{\rm tr}\equiv\theta_{\mathcal N}(T_{\rm QCD})\sim \pi$ until the QCD phase transition with the initial axion velocity $\dot{\theta}_{\rm tr}\equiv\dot{\theta}_{\mathcal N}(T_{\rm QCD})$.
Then the mean velocity is given by 
\begin{eqnarray}
\sqrt{\langle\dot{\theta}_{\rm tr}^2\rangle}=\dfrac{1}{\sqrt {2}} \sqrt{m_{{\mathcal N},1} m_{{\mathcal N},\pi}} |\theta_{1,{\mathcal N}}-\pi| \left(\dfrac{a_{1,{\mathcal N}}}{a_{\rm QCD}}\right)^{3/2}\, .
\end{eqnarray}
At $T=T_{\rm QCD}$, the $Z_{\mathcal N}$ axion mass is exponentially suppressed due to the $Z_{\mathcal N}$ symmetry and the true minimum will develop.
Note that the axion energy density at $T_{\rm QCD}$ is non-adiabatic, which then can be described by
\begin{eqnarray}
\rho_{{\mathcal N}, \rm tr}=\dfrac{1}{2} F_a^2 \dot{\theta}_{\rm tr}^2 + 2\dfrac{m_{\mathcal N}^2 F_a^2}{{\mathcal N}^2}\, .
\end{eqnarray}
The follow-up is determined by the velocity $\dot{\theta}_{\rm tr}\sim 2m_{\mathcal N}/{\mathcal N}$ and we consider a case that the axion has a large initial velocity, $\rm i.e.$, $\dot{\theta}_{\rm tr}\gg 2m_{\mathcal N}/{\mathcal N}$.
Then at $T_2<T<T_{\rm QCD}$, the $Z_{\mathcal N}$ axion energy density is conserved with the PQ charge $q_{\rm kin}=\dot{\theta}_{\mathcal N} a^3$, and here $T_2$ is defined at which the kinetic energy is comparable to the barrier height, $\dot{\theta}_{\mathcal N}(T_2)=2m_{\mathcal N}/{\mathcal N}$.
Using the conserved $q_{\rm kin}$, the scale factor at $T_2$ is given by 
\begin{eqnarray}
a_2=\left(\dfrac{{\mathcal N}\dot{\theta}_{\rm tr}}{2m_{\mathcal N}}\right)^{1/3}a_{\rm QCD}\, .
\end{eqnarray}
At $T<T_2$, the $Z_{\mathcal N}$ axion will start second to oscillate until nowadays, and the axion energy density is adiabatic invariant.
Using 
\begin{eqnarray}
\dfrac{\rho_{{\mathcal N},2} a_2^3}{m_{{\mathcal N},2}}=\dfrac{\rho_{{\mathcal N},0} a_0^3}{m_{{\mathcal N},0}}\, ,
\end{eqnarray}
then the $Z_{\mathcal N}$ axion energy density at present can be described by
\begin{eqnarray}
\rho_{{\mathcal N},0} = \dfrac{m_{{\mathcal N},0} F_a^2 \dot{\theta}_{\rm tr}}{\mathcal N} \left(\dfrac{a_{\rm QCD}}{a_0}\right)^3\, .
\end{eqnarray}
Substituting the mean velocity, we obtain the axion energy density
\begin{eqnarray}
\rho_{{\mathcal N},0}=C \dfrac{m_{{\mathcal N},0} \sqrt{m_{{\mathcal N},1} m_{{\mathcal N},\pi}} |\theta_{1,{\mathcal N}}-\pi| F_a^2}{\sqrt{2}{\mathcal N}}\left(\dfrac{\sqrt{a_{1,{\mathcal N}} a_{\rm QCD}}}{a_0}\right)^3\, ,
\end{eqnarray}
where $C\simeq2$ is a constant.

\medskip\noindent{\bf $Z_{\mathcal N}$ axion in the Case~I.}---%
The axion relic density ratio $R_{\rho}$ in this case is given by
\begin{eqnarray}
\begin{aligned}
R_\rho&\simeq \dfrac{\sqrt{m_{a,0} m_{a,1}} \theta_{1,a}}{\sqrt{m_{{\mathcal N},1} m_{{\mathcal N},\pi}} |\theta_{1,{\mathcal N}}-\pi| \zeta} \left(\dfrac{\sqrt{a_{1,a}}}{\sqrt{a_{1,{\mathcal N}}}}\right)^3\\
&= \sqrt[4]{\dfrac{1-z}{1+z}} \sqrt{\dfrac{m_{a,1}}{m_{{\mathcal N},1}}} \dfrac{\theta_{1,a}}{ |\theta_{1,{\mathcal N}}-\pi| \sqrt{\zeta}} \left(\dfrac{\sqrt{a_{1,a}}}{\sqrt{a_{1,{\mathcal N}}}}\right)^3\\
&= \sqrt[4]{\dfrac{1-z}{1+z}} \sqrt{\dfrac{m_{a,1}}{m_{{\mathcal N},1}}} \dfrac{\theta_{1,a}}{ |\theta_{1,{\mathcal N}}-\pi| \sqrt{\zeta}} \left(\dfrac{m_{{\mathcal N},1}}{m_{a,1}}\right)^{3/4}\\
&= \sqrt[4]{\dfrac{1-z}{1+z}} \sqrt[4]{\dfrac{m_{{\mathcal N},1}}{m_{a,1}}} \dfrac{\theta_{1,a}}{ |\theta_{1,{\mathcal N}}-\pi| \sqrt{\zeta}}\, .
\end{aligned}
\end{eqnarray}

\medskip\noindent{\bf $Z_{\mathcal N}$ axion in the Case~II.}---%
The axion relic density ratio $R_{\rho}$ in this case is given by
\begin{eqnarray}
\begin{aligned}
R_\rho&\simeq \dfrac{\sqrt{m_{{\mathcal N},\pi} m_{a,1}} \theta_{1,a}}{\sqrt{m_{{\mathcal N},1} m_{{\mathcal N},\pi}} |\theta_{1,{\mathcal N}}-\pi| \zeta} \left(\dfrac{\sqrt{a_{1,a}}}{\sqrt{a_{1,{\mathcal N}}}}\right)^3\\
&= \sqrt{\dfrac{m_{a,1}}{m_{{\mathcal N},1}}} \dfrac{\theta_{1,a}}{ |\theta_{1,{\mathcal N}}-\pi| \zeta} \left(\dfrac{\sqrt{a_{1,a}}}{\sqrt{a_{1,{\mathcal N}}}}\right)^3\\
&= \sqrt[4]{\dfrac{m_{{\mathcal N},1}}{m_{a,1}}} \dfrac{\theta_{1,a}}{ |\theta_{1,{\mathcal N}}-\pi| \zeta}\, .
\end{aligned}
\end{eqnarray}

\medskip\noindent{\bf QCD axion in the Case~II.}---%
The axion relic density ratio $R_{\rho}$ in this case is given by
\begin{eqnarray}
\begin{aligned}
R_\rho&\simeq \dfrac{m_{{\mathcal N},1} F_a^2 \theta_{1,{\mathcal N}}^2}{m_{a,1} f_a^2 \theta_{1,a}^2 } \left(\dfrac{a_{1,{\mathcal N}}}{a_{1,a}}\right)^3\\
&= \sqrt{\dfrac{m_{a,1}}{m_{{\mathcal N},1}}} \dfrac{\theta_{1,{\mathcal N}}^2 \zeta^2}{\theta_{1,a}^2}\, .
\end{aligned}
\end{eqnarray}

\end{document}